\documentclass{elsart1p}
\usepackage{graphicx}
\usepackage{amssymb}
\begin{document}
\begin{frontmatter}
%
%
%
\title{Confinement, Turbulence and Diffraction Catastrophes}
%
%
\author[ipht]{J.-P. Blaizot}
and 
\author[smol]{M.A. Nowak}
\address[ipht]{IPhT, CEA-Saclay, 91191 Gif-sur-Yvette, France}
\address[smol]{M. Smoluchowski Institute of Physics,
Jagiellonian University, PL--30--059 Cracow, Poland}
\begin{abstract}
Many features of the large $N_c$ transition that occurs in the
spectral density of Wilson loops as a function of loop area
(observed recently in numerical simulations of Yang-Mills theory by Narayanan and Neuberger)
can be captured by a simple Burgers equation used to model
turbulence. 
 Spectral shock waves that
precede this asymptotic limit exhibit universal scaling with $N_ c$,
with indices that can be related to Berry indices for 
diffraction catastrophes.

\end{abstract}
\end{frontmatter}
%
\section{Spectral shock waves}
Recent lattice studies~\cite{NN} have provided evidence for a novel
universal phenomenon in Yang-Mills
theory with a large number of colors $N$. For small loops, the eigenvalues of the unitary  Wilson operator
concentrate  around the point $z=1$ in the complex plane, and
spread symmetrically on the unit circle towards $z=-1$ as the area of the loop
increases. At some critical area, the spectral gap closes, and the
corresponding level density is given by a universal
function (Pearcey function), whose two arguments exhibit
 scalings in $N^{1/2}$ and $N^{3/4}$, respectively, apparently independently on
the dimensionality of the Yang-Mills theory. For very large loops, the level density  goes towards a universal distribution, reflecting disorder, as
suggested long ago by Durhuus and Olesen~\cite{DO}.  In a recent
paper~\cite{BNPRL}, we suggested a particular mechanism leading to
such transition, by noting that in $d=2$, the
spectral evolution of Wilson loops is governed by a complex Burgers equation \begin{eqnarray}
\label{holoB}
\partial_A F +F\partial_{\theta}F =0, \qquad F(\theta,A)=\frac{1}{2}\int_{-\pi}^{\pi}d\alpha\,
\rho(\alpha,A) \cot ((\theta-\alpha)/{2})
\end{eqnarray}
where the imaginary part of the resolvent $F(\theta,A)$ gives
spectral density $\rho$.  All known properties of the Durhuus and Olesen transition  can be easily recovered by analyzing the
solution of Eq.~(\ref{holoB}), using the methods of complex
characteristics, and following the motion of singularities in the
complex plane \cite{BNPRL}. In  this description, the
transition corresponds to the  collision of two``tsunami-like" shock
waves (associated to the edges of the spectrum) evolving
symmetrically from $z=1$ around the unit circle, to merge violently
at $z=-1$ at the critical area. For larger areas, the cascade of
Fourier modes evolves in such a way that only the lowest mode,
corresponding to the size of the system, survives, a phenomenon
reminiscent of the inverse turbulent cascade.
%
%
 \section{Non-linear diffusion, universality and colored catastrophes}
The complex Burgers equation
describes, under certain conditions, the random  walk of eigenvalues
of large unitary matrices~\cite{DYSON}.The role of the time $t$  is played by the
area of the loop. Writing  the eigenvalues as $z=e^{i\theta}$, the
Langevin equation corresponding to the random walk of unitary
matrices reads
\begin{eqnarray}
d\theta_i =dB_i+\frac{1}{2} \sum_{i \neq j} \cot \frac{\theta_i 
-\theta_j}{2}\,dt
\end{eqnarray}
where the first term is a Brownian noise and the second is 
originating from the usual repulsion of eigenvalues. From the Langevin equation,
a Smoluchowski-Fokker-Planck equation is obtained using standard techniques, which,
after a suitable approximation yields the following equation for $\rho$ \cite{BN09}
 \begin{eqnarray}\label{viscidBurgers}
\partial_t \rho + \partial_{\theta}\left[ \rho {\rm H}[\rho]
\right]=\frac{1}{2}\partial_{\theta \theta} \rho,\qquad
H[\rho]\equiv \frac{1}{2\pi}{\rm {P.V.}} \int_{-\pi}^{\pi}d\alpha
\rho(\alpha,A) \cot\frac{\theta-\alpha}{2},
\end{eqnarray}
where ${\rm {P.V.}}$ denotes the principal value of the integral.
Eq.~(\ref{viscidBurgers}) is equivalent to the viscid Burgers
equation, the last term playing the role of a viscosity term.
This equation exhibits  various scaling with $N$, which we now discuss.\\
$\bullet$ {\it Dyson gas}: We rescale  the density, originally normalized to
$N$, as $\rho=N\hat{\rho}$, and  the time as $\tau=N t$. Then, in the  large $N$ limit we get ~\cite{BS} (equivalent to Eq.~\ref{holoB}) \begin{eqnarray}
\partial_{\tau}\hat{\rho} + \partial_{\theta}
[\hat{\rho}H[\hat{\rho}]]=0.
\end{eqnarray}
Note that the viscous term has
disappeared, since it was dwarfed by the effective spectral viscosity, $v_s=1/2N$ \footnote{
This value of the spectral viscosity has also been obtained by
Neuberger~\cite{N} using different arguments.}. In this regime the evolution proceeds
without collisions.\\
$\bullet$ {\it Dyson liquid}: There is however another regime,
where collisions between the eigenvalues, taken into account by the viscous term,  cannot be ignored. This happens in the vicinity of the shock waves, and seems to be
responsible for the  universal scaling at the edges of
the spectrum. To approach this limit, we set
$
\rho(\theta,t)=N^{\alpha+1}\tilde \rho(x,\tau)$ with $ x\equiv N^\alpha\theta$ and $ \tau\equiv N^{2\alpha}t$, and get
\begin{eqnarray}
\partial_\tau \tilde\rho(x,\tau) + N \partial_x (\tilde\rho H \tilde\rho)=\frac{1}{2}\partial_{xx} \tilde\rho.
\end{eqnarray}
This viscid complex Burgers equation allows us to describe the scaling properties
of the spectral density in the vicinity of a shock.

This can be seen schematically by applying the Cole-Hopf trick to solve the equation~\cite{CL}. The resulting
spectral distribution can be written in the form
\begin{eqnarray}
\tilde\rho(x)= -
 \frac{1}{N \pi}
 {\rm Im}
 \,\partial_x \ln K(x)
\end{eqnarray}
 where $K(x)$ is
the convolution of a Gaussian kernel with $K_0(z)=\exp \{ -N \int^z
\tilde F_0(w)dw\}$, where $\tilde F_0$ is the resolvent
corresponding to the appropriate initial condition for the
microscopic spectral density $\tilde \rho$. In the vicinity of the
shock waves, the resolvent can be approximated by a polynomial of
second order (gapped phase) or third order (at the closure of the
gap). The resulting
spectral densities are therefore related to complex Airy or Pearcey
type integrals, with well defined scaling properties.

In fact these scaling properties appear to be universal and manifest themselves
in various physical contexts. We mention here an amusing analogy with optics.
In geometric optics (vanishing wavelength $\lambda$), the rays of light may condense on
surfaces of infinite intensity, the so-called caustics. These correspond to stable
singularities that are classified, the lowest two corresponding to fold and
cusp singularities. At finite wavelength, interference
effects enter, and we have to use wave packets to describe the
emergence of singularities. Such a wave packet $\Psi$
is generically of the form~\cite{BERRY} \begin{eqnarray}
\Psi(\vec{r},\lambda)=\frac{1}{\lambda^{\beta}}
\psi(\frac{x_1}{\lambda^{\sigma_1}},
\frac{x_2}{\lambda^{\sigma_2}},..)
\end{eqnarray}
For the fold, $\beta=1/6$ and  $\sigma=2/3$, and $\Psi$ is given by an
Airy function, while for the cusp $\beta=1/4$, 
$\sigma_1=1/2, \sigma_2=3/4$, and $\Psi$ is given by a Pearcey
function~\cite{BERRY}. The analogy with these ``diffractive
catastrophes'' goes as follows. The limit of geometrical optics corresponds to the
limit $N=\infty$ ($\lambda\sim 1/N$). Rays of light correspond to the
(complex) characteristics of the Burgers equation, whose envelopes correspond to the caustics and are associated
to singularities of Burgers equation. The universal behavior in the
vicinity of these singularities is captured by a  Pearcey function, and the particular critical exponents for
scaling with ``angle" and ``area" observed in~\cite{NN} are identical to the  Berry indices for a cusp catastrophe.

This work was supported by Marie Curie TOK Grant MTKD-CT-517186
``Correlations in Complex Systems" (COCOS). 

\end{document}